\documentclass[12pt]{article}
\usepackage{amssymb}

\begin{document}

\author{\textbf{G. Baur$^{(a)}$, C.A. Bertulani$^{(b)}$ and D. Dolci$^{(b)}$}\\
\textrm{\small $(a)$ IKP, Forschunszentrum Juelich, Postfach 1913, D-52425
Juelich, Germany \footnote {e-mail: G.Baur@fz-juelich.de }} \\
\textrm{{\small $(b)$ Instituto de F\'{\i }sica, Universidade Federal do Rio
de Janeiro}}\\
\textrm{\small 21945-970 Rio de Janeiro, RJ, Brazil 
\footnote {e-mails: bertu@if.ufrj.br and dolci@if.ufrj.br}}}

\title{Influence of damping on the excitation of the double giant resonance}
\date{}
\maketitle

\begin{abstract}
We study the effect of the spreading widths on the excitation probabilities
of the double giant dipole resonance. We solve the coupled-channels
equations for the excitation of the giant dipole resonance and the double
giant dipole resonance. Taking $Pb+Pb$ collisions as example, we study the
resulting effect on the excitation amplitudes, and cross sections as a
function of the width of the states and of the bombarding energy.
\end{abstract}

Double giant dipole resonances have been mainly studied in heavy ion Coulomb
excitation experiments at high energies (for a recent review, see \cite
{Ber99}). The feasibility of such experiments has been predicted in 1986 
\cite{BB86a,BB86b} where the magnitude of the cross sections for the
excitation of the Double Giant Dipole Resonance (DGDR) was calculated
(see also \cite{Br85}). In
ref. \cite{BB86b} a recipe was given for treating the effect of the width of
the giant resonances on the excitation probabilities and cross sections. In
this letter we make a quantitative prediction of this effect using a
realistic coupled-channels calculation for the excitation amplitudes. The
coupling interaction for the nuclear excitation $i\longrightarrow f$ in a
semiclassical calculation for a electric $\left( \pi =E\right) $, or
magnetic $\left( \pi =M\right) $, multipolarity, is given by (eqs. (6-7) of
ref. \cite{Ber98})

\begin{equation}
W_C=\frac{V_C}{\epsilon _0}=\sum_{\pi \lambda \mu }W_{\pi \lambda \mu
}\left( \tau \right) \;,  \label{wc1}
\end{equation}
where

\begin{equation}
W_{\pi \lambda \mu }\left( \tau \right) =\left( -1\right) ^{\lambda +1}\frac{%
Z_1e}{\hbar vb^\lambda }\ \frac 1\lambda\ \sqrt{\frac{2\pi }{\left( 2\lambda
+1\right) !!}}Q_{\pi \lambda \mu }\left( \xi ,\tau \right) \mathcal{M}\left(
\pi \lambda ,\mu \right) \;.  \label{wc2}
\end{equation}

Above, $b$ is the impact parameter, $\gamma =\left( 1-\beta ^2\right)
^{-1/2} $, $\beta =v/c$, $\tau =\gamma vt/b$ is a dimensionless time
variable, $\epsilon _0=\gamma \hbar v/b$ sets the energy scale and $Q_{\pi
\lambda \mu }\left( \xi ,\tau \right) $, with $\xi =\xi _{if}=\left(
E_f-E_i\right) /\epsilon _0$ as an adiabatic parameter, depends exclusively
on the properties of the projectile-target relative motion. The multipole
operators, which act on the intrinsic degrees of freedom are, as usual,

\begin{equation}
\mathcal{M}(E\lambda ,\mu )=\int d^3r \ \rho (\mathbf{r})\ r^\lambda \
Y_{1\mu }(\mathbf{r})\ ,  \label{ME1}
\end{equation}
and

\begin{equation}
\mathcal{M}(M1,\mu )=-{\frac{i}{2c}}\ \int d^{3}r\ \mathbf{J}(\mathbf{r}).%
\mathbf{L}\left( rY_{1\mu }\right) \ ,  \label{MM1}
\end{equation}
We treat the excitation problem by the method of Alder and Winther\cite{AW}.
We solve a time-dependent Schr\"{o}dinger equation for the intrinsic degrees
of freedom in which the time dependence arises from the projectile-target
motion, approximated by the classical trajectory. For relativistic energies,
a straight line trajectory is a good approximation. We expand the wave
function in the set $\{\mid k\rangle ;\ k=0,N\}$ of eigenstates of the
nuclear Hamiltonian, where 0 denotes the ground state and $N$ is the number of intrinsic excited states included in
the coupled-channels (CC) problem. We obtain a set of coupled equations.

To simplify the expression we introduce the dimensionless parameter $\Theta
_{kj}^{(\lambda\mu )}$ by the relation 
\begin{equation}
\Theta _{kj}^{(\lambda\mu )}=\left( -1\right) ^{\lambda +1}\frac{Z_{1}e}{\hbar 
\mathrm{v}b^{\lambda }}\frac{1}{\lambda }\sqrt{\frac{2\pi }{\left( 2\lambda
+1\right) !!}}\;\mathcal{M}_{kj}(E\lambda )
\end{equation}

Then we write the coupled channels equations in the form \cite{Ber98} 
\begin{equation}
\frac{da_{k}(\tau )}{d\tau }=-i\sum_{j=0}^{N}\ \sum_{\pi \lambda \mu }Q_{\pi
\lambda \mu }(\xi _{kj},\tau )\Theta _{kj}^{\left( \lambda \mu \right) }\;\exp
\left( i\xi _{kj}\tau \right) \;a_{j}(\tau )\;.  \label{ats1}
\end{equation}

In what follows we concentrate on the $E1$ excitation mode. In this case, we
have

\begin{equation}
Q_{E10}(\xi ,\tau )=\gamma \sqrt{2}\left[ \tau \phi ^3(\tau)-i\xi \left( \frac{%
\mathrm{v}}c\right) ^2\phi (\tau )\right] \;;\;\;\;\;\;Q_{E1\pm 1}(\xi ,\tau
)=\mp \phi ^3(\tau )\;,  \label{QE1}
\end{equation}
where $\phi (\tau )=\left( 1+\tau ^2\right) ^{-1/2}.$

Following ref. \cite{BB86b} the inclusion of damping leads to the
coupled-channels equations with damping terms, i.e.,

\begin{equation}
\frac{da_{k}(\tau )}{d\tau }=-i\sum_{j=0}^{N}\ \sum_{\mu }Q_{E1\mu }(\xi
_{kj},\tau )\Theta _{kj}^{\left( 1\mu \right) }\;\exp \left( i\xi
_{kj}\tau \right) \;a_{j}(\tau )\;-\frac{\Gamma _{k}}{\epsilon _{0}}%
a_{k}(\tau ).  \label{ccd}
\end{equation}
These equations lead to master equations for the occupation probabilities, $%
\widetilde{P}_{j}(\tau )=\left| a_{j}(\tau )\right| ^{2}$, in the form

\begin{equation}
d\widetilde{P}_{k}(\tau )/d\tau =G_{k}(\tau )-L_{k}(\tau )  \label{mast}
\end{equation}
where

\begin{equation}
G_{k}(\tau )=2 \ \Im m\sum_{j}\sum_{\mu }Q_{E1\mu }(\xi _{kj},\tau
)\Theta _{kj}^{\left( 1\mu \right) }\; \exp({i\xi_{kj}\tau})
\;a_{k}(\tau )\;a_{j}^{*}(\tau )
\label{gain}
\end{equation}
and

\begin{equation}
L_{k}(\tau )=-\frac{\Gamma _{k}}{\epsilon _{0}}\widetilde{P}_{k}(\tau )\;.  \label{loss}
\end{equation}

These equations can be integrated, yielding the conservation law,

\begin{equation}
\sum_{k}\Biggl( \widetilde{P}_{k}(\tau )+\widetilde{F}_{k}(\tau
)\Biggr)=1\;,\;\;\;\;\;\;\;\; {\rm where}\;\;\; \widetilde{F}_{k}(\tau )=\;
\frac{\Gamma _{k}}{\epsilon
_{0}}\int_{-\infty }^{\tau }\widetilde{P}_{k}(\tau ^{\prime })d\tau ^{\prime
}  \label{conserv}
\end{equation}

Due to the exponential decay of the states with $k\geqslant 1$, we have for $%
t\rightarrow \infty $ the limit $\widetilde{P}_{k}(\infty )=\delta _{j0}%
\widetilde{P}_{0}(\infty )$ and

\begin{equation}
\widetilde{P}_{0}(\infty )+\sum_{k}\widetilde{F}_{k}(\infty )=1\;.
\end{equation}

This means that for $t\rightarrow \infty $ there is a probability to find
the system in the ground state given by $\widetilde{P}_{0}(\infty )$ and a
probability that it has been excited and decayed through the channel $j$
which is given by $\widetilde{F}_{j}(\infty )$. Thus, the set of equations 
\ref{ccd} are shown to correctly describe the contribution to the
excitation through channel $j$.

The excitation probability of an intrinsic state $\mid j\rangle $ in a
collision with impact parameter $b$ is obtained from an average over the
initial orientation and a sum over the final orientation of the nucleus, as 
\begin{equation}
P_{j}(b)=\frac{1}{2I_{0}+1}\sum_{M_0,\; M_{j}}|\widetilde{F}_{j}(\infty )|^{2}\;,
\label{Pn}
\end{equation}
and the cross section is obtained by the classical expression 
\begin{equation}
\sigma _{j}=2\pi \ \int P_{j}(b)\ T(b)\ b\;db\;.
\end{equation}
Above, $T(b)$ accounts for absorption according to the prescription of ref. 
\cite{Ber98}, using the nucleon-nucleon cross sections and the ground state
density of $Pb$ from experimental data.

We consider the excitation of giant resonances in $^{208}$Pb projectiles,
incident on $^{208}$Pb targets at 640 A$\cdot $MeV. This reaction has been
studied at the GSI/SIS, Darmstadt~\cite{hans}. For this system the
excitation probabilities of the isovector giant dipole ($GDR$) at 13.5 MeV
are large and, consequently, high order effects of channel coupling should
be relevant. To assess the importance of the damping effects, we calculate
the matrix elements assuming that the GDR is an isolated state depleting
100\% of the energy-weighted sum-rule. The matrix element for the GDR$%
\rightarrow $ DGDR transition incorporates the boson factor $\sqrt{2}$, as
usual \cite{BB86b}. The energy location of the DGDR state is taken as 27
MeV, consistent with the experimental data. The spin and parities of the
states are given by $1^{-}$ for the GDR, and $0^{+}$ and $2^{+}$ for the
DGDR, respectively. The distribution of the strength among the  $0^{+}$ and $%
2^{+}$ DGDR states are simply obtained from Clebsh-Gordan coefficients \cite
{BZ93}.

In figure 1 we plot the time-dependent occupation probabilities of the ground state, in $Pb$%
, $N=0$, of the GDR state, $N=1$, and of the DGDR state, $N=2$,
respectively. Figure 1(a) shows the occupation probabilities with the widths
equal to zero, $\Gamma _{N}=0$. In figure 1(b) we plot the occupation
probabilities of the GDR state, $N=1$, and of the DGDR state, $N=2$, with $%
\Gamma _{N}=0$ (full lines), and with $\Gamma _{GDR}=4$ MeV (experimental),
and $\Gamma _{DGDR}=5.7$ MeV (dashed lines). The width of the DGDR is set to $\Gamma
_{DGDR}=\sqrt{2}\Gamma _{GDR}$, following the apparent trend of the experimental
data \cite{hans}. Note, that $\Gamma
_{DGDR}=2\Gamma _{GDR}$ has a better (and simpler) theoretical explanation
\cite{BB86c}. But, the time integrated population of the DGDR state will
not be much influenced by using the later parametrization. 

We observe that the inclusion of damping leads to strong
modifications in the time-dependent occupation probabilities of the GDR and
DGDR states. One might wrongly deduce from figure 1 that the excitation
probabilities are reduced proportionally to the difference between the
maximum value of $\widetilde{P}_{N}(\tau )$, with and without damping.
However, the quantities shown in figure 1(b) includes the loss of occupation
probabilities of a given state, \textit{while} it is being populated by the
time-dependent transitions. Thus, the reduction of the excitation
probabilities of the GDR and the DGDR due to damping is smaller than deduced
from figure 1. The relevant quantity to calculate the excitation cross
sections are the quantities $\widetilde{F}_{j}(\infty )$, which account for
the time-integrated transition probability, followed by decay, of the state $%
j$.

In figure 2 we plot the flux functions, or time-integrated transition
probabilities to the GDR and the DGDR states as a function of the width of
the collective state, $\Gamma _{GDR}$, keeping constant the ratio $\Gamma
_{DGDR}/\Gamma _{GDR}=\sqrt{2}$. We keep the impact parameter fixed, $b=15$
fm. We note that, varying the width from 0 up to 4 MeV leads to a 10\%
decrease of the flux functions into GDR and DGDR states. A similar tendency
is observed for the total cross section, integrated over impact parameters.
The excitation probability decreases with about $1/b^4$, therefore the grazing
collisions are weighted most strongly.

In figure 3(a) we plot the effect of damping in the total cross sections, as
a function of the bombarding energy. We take $\Gamma _{GDR}=0$ (dashed line)
and $\Gamma _{GDR}=4$ MeV (full line), keeping constant the ratio $\Gamma
_{DGDR}/\Gamma _{GDR}=\sqrt{2}$. We observe that the effect of damping
disappears, as the bombarding energy increases. At high energies the
reaction is fast, and the system does not have time to dissipate during its
excitation. In this regime the sudden approximation is a valid approach to
the calculation of the excitation amplitudes. 

We note that our model is restricted to the excitation of isolated resonant
states, including a time-dependent loss term on the far right of the
coupled-channels equations \ref{ccd}. This is different from a study of the
influence of the fragmentation of the resonances into many neighbouring
states. In this case, the Coulomb excitation of the giant resonances is
obtained as a superposition of excitations to states spread over an energy
envelope, usually taken as a Lorentzian shape. States at lower energy are
more easily excited than states at higher energies. Thus, the spreading of
the resonances may lead to another kind of effect of the widths, not
obtainable in the above treatment. To study this effect in a simple way, we
use the harmonic model of ref. \cite{abs95}. The photo-nuclear cross
sections which enter in these calculations are of Lorentzian shapes, with a
cut at the low energy limit of 8 MeV, corresponding to the threshold for
neutron emission, since most of the DGDR manifestation in experiments come
from neutron emission after relativistic Coulomb excitation.
The magnitude of the  photo-nuclear cross sections are 
obtained by using a 100\% depletion of the Thomas-Reiche-Kuhn
energy-weighted sum-rule applied to the GDR. 

The harmonic model provides a
simple analytical formula to calculate the excitation probabilities of the
DGDR \cite{abs95}. The resulting cross sections are 
shown in figure 3(b) where in the solid
line we take $\Gamma _{GDR}=\Gamma _{DGDR}=0$, while the dashed-lines are
for $\Gamma _{DGDR}=5.7$ MeV and $\Gamma _{GDR}=4$ MeV. We observe a similar
effect as in figure 3(a). This is understood as a reduction due to the
spread of states at energies above the energy centroid of the Lorentzian
envelope. The excitation amplitude for these states are smaller, thus
leading to a net reduction of the energy integrated Coulomb excitation cross
sections.

In conclusion, we have obtained the dependence of the excitation amplitudes
on the width of the giant resonance states. We show that the effect reduces
excitation probabilities, and cross sections. We have developed an approach
to solve this problem in realistic situations. It is demonstrated that the
dynamical effect of the widths of the GR's in a time-dependent picture leads
to a decrease of the cross sections, more accentuated for low energy
collisions. The energy fragmentation of the giant resonances can be studied
in a simple fashion within the harmonic model. The net effect is also to
decrease the cross sections with increasing width, specially at low energy
collisions.

\textbf{Acknowledgments}

This work was supported in part by the Brazilian funding agencies CNPq,
FAPERJ, FUJB/UFRJ, and PRONEX, under contract 41.96.0886.00.

\textbf{Figure Captions}

Fig. 1 - Occupation probabilities in Coulomb excitation of $Pb$, in $Pb+Pb$
collisions at 640 MeV.A. $N=0$ for the ground state, $N=1$ for the GDR
state, $N=2$, for the DGDR state, respectively. Figure 1(a) shows the
occupation probabilities with the widths equal to zero, $\Gamma _{N}=0$. In
figure 1(b) we plot the occupation probabilities of the GDR state, $N=1$,
and of the DGDR state, $N=2$, with $\Gamma _{N}=0$ (full lines), and with $%
\Gamma _{GDR}=4$ MeV, and $\Gamma _{DGDR}=5.7$ MeV (dashed lines). The width
of the DGDR is set to $\Gamma _{DGDR}=\sqrt{2}\Gamma _{GDR}$, according to
the trend of the experimental data \cite{hans}.

Fig. 2 - Flux functions, or time-integrated transition probabilities to the
GDR and the DGDR states in $Pb$ as a function of the width of the collective
state, $\Gamma _{GDR}$, keeping constant the ratio $\Gamma _{DGDR}/\Gamma
_{GDR}=\sqrt{2}$. We keep the impact parameter fixed, $b=15$ fm. $N=1$ for
the GDR state, $N=2$, for the DGDR state.

Fig. 3 - (a) Total cross sections, as a function of the bombarding energy. We
take $\Gamma _{GDR}=0$ (dashed line) and $\Gamma _{GDR}=4$ MeV (full line),
keeping constant the ratio $\Gamma _{DGDR}/\Gamma _{GDR}=\sqrt{2}$. $N=1$
for the GDR state, $N=2$, for the DGDR state. (b) Results of the harmonic
model \cite{abs95} for the Coulomb excitation cross sections of the GDR ($N=1
$) and the DGDR ($N=2$). The solid lines use $\Gamma _{GDR}=\Gamma _{DGDR}=0$%
, while the dashed-lines are for $\Gamma _{DGDR}=5.7$ MeV and $\Gamma
_{GDR}=4$ MeV.

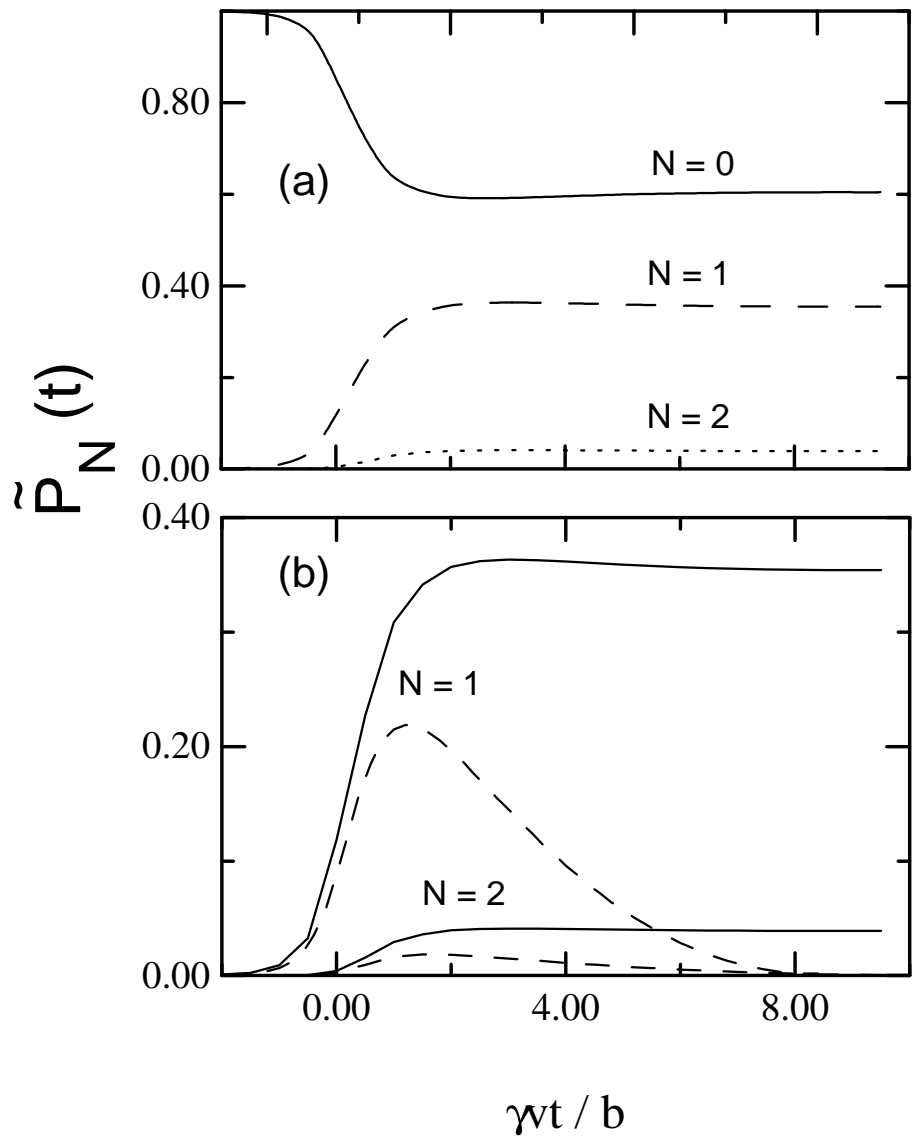
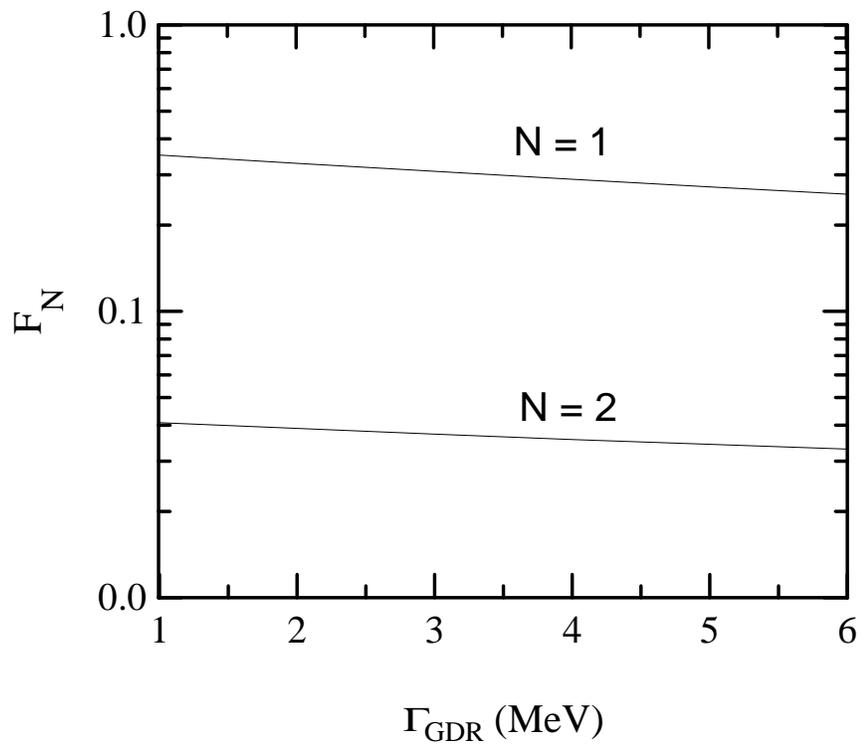
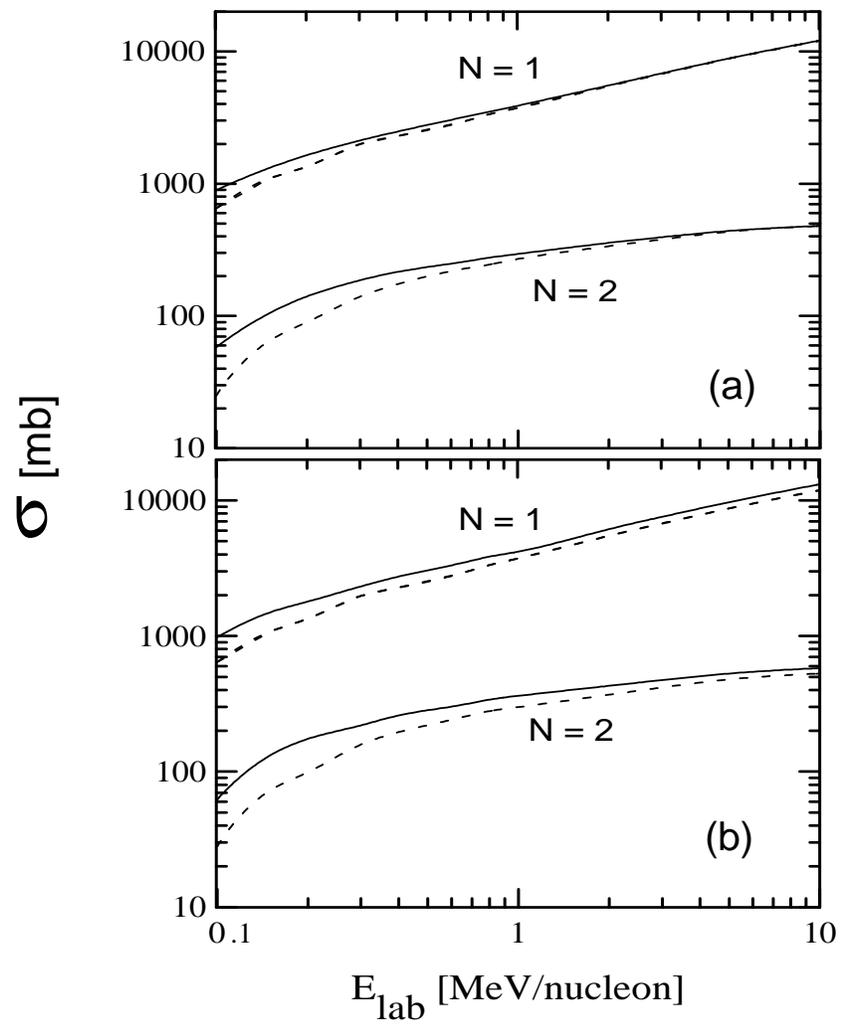
\end{document}